\documentstyle[11pt]{article}
\include{epsf}
\begin{document}

\thispagestyle{empty}

\begin{flushright}
MIT-CTP-2800 \\
hep-lat/9811014
\end{flushright}
\begin{center}
\vspace*{10mm}
{\LARGE Loop representation for 2-D Wilson lattice    
\vskip2mm
fermions in a scalar background field} 
\vskip18mm
\centerline{ {\bf
Christof Gattringer}}
\vskip 2mm
Massachusetts Institute of Technology \\
Center for Theoretical Physics \\
77 Massachusetts Avenue \\
Cambridge MA 02129, USA
\vskip30mm
\begin{abstract}
We show that the fermion determinant for 2-D Wilson 
lattice fer\-mi\-ons coupled
to an external scalar field is equivalent to self avoiding loops interacting
with the external field. In an application of the resulting formula
we integrate the scalar field with a Gaussian action
to generate the $N$-component Gross-Neveu model. The loop representation for 
this model is discussed. 
\end{abstract}
\end{center}
\newpage
\setcounter{page}{1}
\section{Introduction} 
\noindent
The fermion determinant is a highly non-local object 
as can e.g.~be seen from the hopping expansion for lattice regularized 
fermions (see \cite{MoMu94} for a basic introduction). 
The hopping expansion expresses the fermion determinant as the exponential
of a sum over all possible closed loops and the external 
field variables along a loop are collected as factors for this
loop. The exponential function can then be expanded and the result is a 
representation of the fermion determinant in terms of loops.
The loop-sum in the exponent, however,
contains loops of arbitrary length, which can also iterate parts or all of
their contour arbitrarily often. Thus, the exponent contains arbitrarily
high powers of the external fields. On the other hand we know that, at least on
a finite lattice, the fermion determinant is a finite polynomial in the 
external fields, and thus infinitely many contributions present in the
exponent have to cancel each other when expanding the exponential.
The result has to be a relatively simple loop representation for the
fermion determinant. 

For the case of staggered fermions several papers can be found in the 
literature \cite{RoWo84}-\cite{Mo90} where polymer representations
for the fermions are obtained. For the case of Wilson fermions relatively
few is known due to the more involved spinor structure of the fermions. 
Here we concentrate on 2-D Wilson fermions.

An instance where the above discussed cancellation of contributions was 
brought under control for a model with Wilson fermions is Salmhofer's 
mapping of the strongly coupled lattice 
Schwinger model to a self avoiding loop model \cite{Sa91}. In a first
step the gauge fields at strong (=infinite) coupling were integrated 
out and the remaining Grassmann integral was then represented as a
sum over loops. Subsequently Scharnhorst studied the two-flavor lattice
Schwinger model with this method \cite{Sch96} and extended the techniques
to find a two-color loop model for the 2-D lattice Thirring model 
\cite{Sch97}. In all of these cases, however the external field was either 
integrated out in the 
strong coupling limit \cite{Sa91,Sch96} or wasn't present at all 
\cite{Sch97}. On the other hand, our arguments given in the first 
paragraph show that the cancellation of higher winding loops is a 
universal phenomenon and that it should be possible to find a simple 
loop representation also for the fermion determinant in an external field. 

In this article we present a simple formula for the fermion determinant
of 2-D Wilson lattice fermions in a scalar background field
(Eq.~(\ref{equivalence})). The proof
is based on a result for the hopping expansion for a generalized 8-vertex
model where the vertices are coupled to a background field \cite{Ga98b}. 
With a proper choice of the vertex weights the square of the partition
function for this model represents
the 2-D fermion determinant in a scalar background field. The resulting
expression reduces the hopping expansion to a finite sum (on a finite lattice)
of loops. In addition it is possible to explicitly integrate out
the external field. By doing so with a Gaussian action for the scalar
we generate a loop representation of the 2-D Gross-Neveu model. Such loop
representations for lattice field theories allow for a considerably more 
accurate
numerical treatment as has e.g.~been demonstrated for the strongly coupled 
Schwinger model \cite{GaLaSa92}.

\section{Setting and hopping expansion } 
\noindent
The basic idea for the proof of our loop representation is to identify the 
hopping expansion of the Wilson fermion determinant with the hopping
expansion of a generalized 8-vertex model \cite{Ga98b}. Here we briefly
rederive the hopping expansion for Wilson fermions in a form suitable
for comparison with the generalized 8-vertex model
(see e.g.~\cite{MoMu94} for a more 
detailed discussion of the hopping expansion).  

We study a lattice model of 2-D fermions which in the continuum corresponds
to the action
\begin{equation}
S \; = \; \int d^2 x \; \overline{\psi}(x) \;
\Big[ \gamma_\mu \partial_\mu \; - \; m 
\; - \; \theta(x) \Big] \; \psi(x) \; ,
\label{continuum}
\end{equation}
where $\theta(x)$ is a scalar external field. 
The lattice fermion determinant is expressed as a path integral
\begin{equation}
\det M[\theta] \; = \; \int D\psi D\overline{\psi} \; 
\exp\left(-\sum_{x,y \in \Lambda} \overline{\psi}(x) 
M[\theta](x,y) \psi(y) \right) \; ,
\label{partfunct}
\end{equation}
where the kernel for the lattice fermion action (Wilson fermions)
which regularizes the continuum action (\ref{continuum}) is given by
\[
M[\theta](x,y) \; = \; \Big[2 + m + \theta(x)\Big] \; \delta_{x,y} 
\; - \; \sum_{\mu = \pm 1}^{\pm 2} \Gamma_\mu \delta_{x+\mu,y} \; ,
\]
and we defined
\[
\Gamma_{\pm \mu} \; = \; \frac{1}{2} [ 1 \mp \sigma_\mu ]
\; \; \; \; , \; \; \; \; \mu = 1,2 \; .
\]
Here $\sigma_1,\sigma_2$ are Pauli matrices. The sum in the exponent 
of (\ref{partfunct}) runs over the whole lattice $\Lambda$, 
which for simplicity we 
assume to be a finite rectangular piece of Z\hspace{-1.3mm}Z$^2$
(the generalization to e.g.~a torus is straightforward).
The boundary conditions are open, i.e.~hopping terms that would lead to the 
outside of our lattice are omitted. We define
\[
h(x) \; = \; 2 + m + \theta(x) \; ,
\]
and assume that $\theta(x)$ is such that $h(x) \neq 0$ for all lattice points
$x$. This is a purely technical assumption due to the particular techniques 
we use for computing the determinant. The final result will be a finite 
polynomial in the $\theta(x)$ and the above restriction is irrelevant then.
We now can write
\begin{equation}
\det M[\theta] \; = \; \prod_{x \in \Lambda} h(x)^2 
\det\Big(1 - R[\theta] \Big)
\; = \; \prod_{x \in \Lambda} h(x)^2
\exp \left( - \sum_{n=1}^\infty \frac{1}{n} \mbox{Tr} \; 
R[\theta]^n \right) \; .
\label{hopexp}
\end{equation}
In the last step the hopping expansion was performed, i.e. the determinant 
was expressed using the well known trace-logarithm formula and the logarithm
was expanded in a power series.
The hopping matrix $R[\theta]$ is defined as
\begin{equation}
R[\theta](x,y) \; = \; \sum_{\mu = \pm 1}^{\pm 2} \;
\Gamma_{\mu} \; \frac{1}{h(x)} \; \delta_{x+\mu, y} \; .
\label{hoppingmatrix}
\end{equation}
The series in the exponent of (\ref{hopexp}) converges for 
$||R[\theta]|| < 1$, which can be enforced by choosing large enough
$h(x)$, i.e.~suitable $\theta(x)$. Again this is only a 
technical restriction and can be abandoned
in the final result. 
Due to the Kronecker delta in (\ref{hoppingmatrix}),
the contributions to Tr$R[\theta]^n$ are supported 
on closed loops on the lattice, and since closed loops are of even length,
the contributions for odd $n$ vanish. For even $n = 2k$ we obtain
\begin{equation}
\mbox{Tr} \; R[\theta]^{2k} \; = \; 
\sum_{x \in \Lambda} \; \sum_{l \in {\cal L}^{(2k)}_x} \; \prod_{y \in P(l)}
\frac{1}{h(y)} \; \; \mbox{Tr} \; \prod_{\mu \in l} \Gamma_\mu \; .
\label{hoptrace}
\end{equation}
Here ${\cal L}^{(2k)}_x$ is the set of all closed, connected loops of length 
$2k$ and base point $x$. By $P(l)$ we denote the set of all sites visited by 
the loop $l$. Note that a factor $1/h(x)$ is produced        
whenever $l$ runs through $x$
which can be arbitrary often for long enough loops. 
The last term in (\ref{hoptrace}) is the
trace of the ordered product of the hopping generators $\Gamma_\mu$ as they 
appear along the loop $l$. We remark, that 
$\Gamma_{\pm \mu} \Gamma_{\mp \mu} = 0$, which implies that whenever
a loop turns around at a site and runs back along its last link
this contribution vanishes. Thus all these {\it back-tracking} loops can be 
excluded from ${\cal L}_x^{(2k)}$.

Evaluating the trace over the matrices $\Gamma_\mu$  for a given loop is the 
remaining problem in the hopping expansion. 
It first has been solved in \cite{St81} by realizing that the Pauli matrices 
give rise to a representation of discrete 
rotations on the lattice. Alternatively one
can decompose the loop using four basic steps and compute the trace in 
an inductive procedure along the lines of \cite{GaJaSe98}. The result is
\begin{equation}
\mbox{Tr} \; \prod_{\mu \in l} \Gamma_{\mu} \; = \; 
-(-1)^{s(l)} \Big( \frac{1}{\sqrt{2}} \Big)^{c(l)} \; .
\label{gtrace}
\end{equation}
By $s(l)$ we denote the number of self-intersections of the loop $l$ and 
$c(l)$ gives its number of corners. The result is independent of the
orientation of the loop.
Inserting (\ref{hoptrace}) and (\ref{gtrace}) in (\ref{hopexp}) we obtain
\begin{equation}
\det M[\theta] \; =  \; \prod_{x \in \Lambda} h(x)^2
\exp \left(  \sum_{k=1}^\infty \frac{1}{2k} 
\sum_{y \in \Lambda} \; \sum_{l \in {\cal L}^{(2k)}_y} 
(-1)^{s(l)} \Big(\frac{1}{\sqrt{2}}\Big)^{c(l)}  
\prod_{z \in P(l)}
\frac{1}{h(z)} \right) .
\label{prelim}
\end{equation}
Finally we further simplify this expression by removing the explicit summation
over the base points $x$. A loop of length $2k$ without complete iteration of 
its contour allows for $2k$ different choices of a base point thus cancelling
the factor $1/2k$ in (\ref{prelim}). A loop which iterates its whole contour
$I(l) > 1$ times allows only for $2k/I(l)$ different base points and a factor
$1/I(l)$ remains. The final expression is
\begin{equation}
\det M[\theta] \; =  \; \prod_{x \in \Lambda} h(x)^2 \;
\exp \left( 2 \sum_{l \in {\cal L}} 
\frac{(-1)^{s(l)}}{I(l)} \Big(\frac{1}{\sqrt{2}}\Big)^{c(l)} \; 
\prod_{y \in P(l)}
\frac{1}{h(y)} \right) \; ,
\label{finalhop}
\end{equation}
where ${\cal L}$ is the set of all closed, connected, non back-tracking 
loops of arbitrary length. Each loop is included in ${\cal L}$ with only one
of its two possible orientations and we collect an overall factor of 2 in
the exponent.

\section{Identification of the corresponding generalized 8-vertex model}
\noindent
As already outlined, the next step is to compare the result (\ref{finalhop})
to the hopping expansion for a generalized 8-vertex model studied in detail 
in \cite{Ga98b}. In this generalized model, the vertices are coupled 
to a locally varying external field $\varphi(x)$. Before we proceed let's
first discuss this generalized model and its relation to the standard 
8-vertex model.

\begin{figure}[htp]
\centerline{\hspace*{-2mm}
\epsfysize=4cm \epsfbox[ 0 0 256 155 ] {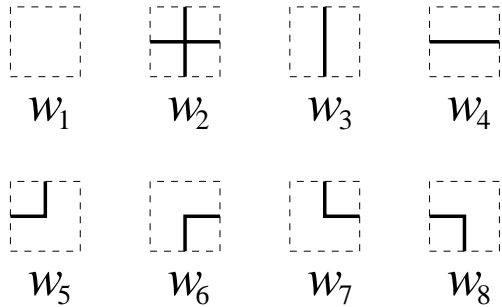}}
\caption{ {\sl The eight vertices (tiles) and their weights $w_i$.}
\label{tiles}}
\end{figure}

The standard 8-vertex model 
\cite{FaWu69,FaWu70,Ba82} can be viewed 
as a model of 8 quadratic tiles (vertices) and 
each of them is assigned a weight $w_i\;  (i = 1, ... 8)$ 
(compare Fig.~\ref{tiles}).
A {\it tiling} of our lattice $\Lambda$ (the same rectangular piece of 
$\mbox{Z\hspace{-1.3mm}Z}^2$ as before)
is a covering of $\Lambda$ with the tiles such that on each site of $\Lambda$
we place one of our tiles with the centers of the tiles sitting on the sites.
The set ${\cal T}$ of {\it admissible tilings} 
is given by those arrangements of 
tiles where the black lines on the tiles never have an open end. The 
partition function of the standard 8-vertex model is the sum over all 
admissible tilings $t \in {\cal T}$
and the Boltzmann weight for a particular tiling $t$  
is given by the product of the weights $w_i$ for all tiles used in 
this tiling $t$.

In our generalization of the model we now couple the vertices to an external
field $\varphi(x)$ located on the sites $x$ of $\Lambda$.
The partition function of the generalized model is given by
\begin{equation}
Z_{8v}[\varphi] \; = \; 
\sum_{t \in {\cal T}} \; \prod_{i = 1}^8 {w_i}^{n_i(t)} \!
\prod_{x \in P(t)} \varphi(x) \; .
\label{8vertex}
\end{equation}
Here $P(t)$ denotes the set of all sites occupied by the tiling $t$.
When a site $x$ is occupied by tile Nr.~2, this site is counted twice
giving a factor $\varphi(x)^2$. In case $x$ is occupied by tile Nr.~1,
$x \notin P(t)$ and the factor is 1. For all other tiles $x$ is 
counted once and the factor is $\varphi(x)$.

The generalized model (\ref{8vertex}) also allows for a hopping expansion
which is derived in \cite{Ga98b}. The central step is to rewrite the partition 
function as an integral over Grassmann variables along the lines of
\cite{Sa80,FrSrSu80,It82}. The action is a quadratic form in the Grassmann
variables and the partition
function gives rise to a Pfaffian. The Pfaffian however can be expanded 
similar to the expansion in Section 2 and it is furthermore possible
to explicitly evaluate the traces over the hopping matrix. The resulting 
expression reads \cite{Ga98b}
\begin{equation}
Z_{8v}[\varphi] \; = \; 
(-1)^{|\Lambda|} w_1^{|\Lambda|} \; 
\exp \left( \sum_{l \in {\cal L}} \frac{(-1)^{s(l)}}{I(l)} 
\Big( \frac{1}{w_1} \Big)^{|l|}
\prod_{i=3}^8 w_i^{n_i(l)} \;
\prod_{x \in P(l)} \varphi(x) 
 \right).
\label{8vertexhop}
\end{equation}
As for the Wilson fermions the sum runs over the set ${\cal L}$ of all closed, 
non back-tracking loops of arbitrary length. Each loop $l$ is included in 
${\cal L}$ with only one of its two possible orientations. By $|l|$ we 
denote the length of the loop, $I(l)$ is the number of 
iterations of its complete contour and $s(l)$ is the number 
of self-intersections. $P(l)$ again denotes the set of sites
visited by the loop $l$ with $x$ being included in $P(l)$ whenever the loop 
$l$ runs through $x$. By $|\Lambda|$ we denote the size of the lattice. 
In the following we will work on lattices with an even number of sites and
the overall sign factor is 1.

The exponents $n_i(l)$ in (\ref{8vertexhop}) give the numbers for the 
abundance of the line 
elements as they are depicted in Fig.~\ref{tiles}. E.g.~when the loop $l$
changes from heading east to heading north at a site it picks up a 
factor of $w_5$ and similarly for the other tiles $w_3, w_4, w_6, w_7, w_8$.
Note that the loops $l \in {\cal L}$ occur as an ordered set of instructions
for the directions the loop takes as it hops from one site to the next. The
element corresponding to the tile 2 with weight $w_2$ 
(compare Fig.~\ref{tiles}) is not needed to describe the loop $l$. Equation
(\ref{8vertexhop}) does not contain $w_2$ explicitly at all. The
weight $w_2$ is related to the other weights through the free fermion 
condition \cite{FaWu69}
\begin{equation}
\omega_1 \omega_2 \; + \; \omega_3 \omega_4 \; = \; 
\omega_5 \omega_6 \; + \; \omega_7 \omega_8 \; .
\label{freecond}
\end{equation}
The free fermion condition is a sufficient condition for finding an explicit
solution to the standard 8-vertex model without external fields. The above 
mentioned Grassmann representation automatically enforces the free fermion 
condition as is discussed in more detail in $\cite{Ga98b}$.

By comparing (\ref{finalhop}) with (\ref{8vertexhop}) we find that the sums
over the loops in the exponent become identical when setting
\[
w_1 = w_3 = w_4 = 1 \; \; \; , \; \; \; w_5 = w_6 = w_7 = w_8 = 
\frac{1}{\sqrt{2}} \; \; \; , \; \; \; \varphi(x) = \frac{1}{h(x)} \; .
\]
Using the free fermion condition (\ref{freecond}) we find $w_2 =0$.  
This implies that when representing Wilson fermions using 
Eq.~(\ref{8vertex}), the sum over 
the tilings can be replaced by a sum over the set ${\cal L}_{sa}$ of 
closed, self-avoiding loops, i.e.~loops that are not allowed to self intersect 
or touch each other. The loops can have several disconnected but closed pieces
and each piece is included in ${\cal L}_{sa}$ with only one of its two possible
orientations.

It is important to note, that the exponents in (\ref{finalhop}) and 
(\ref{8vertexhop}) differ by an overall factor of 2. This is due to the fact
that the action for the Wilson fermions is a bilinear form giving
rise to a determinant while the Grassmann action for (\ref{8vertex}) is a
quadratic form giving rise to a Pfaffian when integrating out the Grassmann
variables. Since the kernel of the latter action is anti-symmetric the
Pfaffian, however, is given by the square root of a determinant causing
the difference by a factor of 2 in the exponent. 

Putting things together we obtain the final formula 
\begin{equation}
\det M[\theta] \; = \; \prod_{x \in \Lambda} [2 + m + \theta(x)]^2 \;
\left( \sum_{l \in {\cal L}_{sa}} \Big( \frac{1}{\sqrt{2}} \Big)^{c(l)}
\prod_{y \in P(l)} \frac{1}{2 + m + \theta(y)} \right)^2 \; ,
\label{equivalence}
\end{equation}
where as above $c(l)$ is the number of corners of $l$. The loops in 
${\cal L}_{sa}$ are self-avoiding and thus for a given loop configuration
$l$ each site of the lattice is occupied only once. Thus in the sum each 
inverse field $[2 + m + \theta(y)]^{-1}$ can only occur linearly. When 
taking the square of this sum 
we can only produce terms which are at most quadratic in 
the inverse field. The overall factor, however, still can cancel the quadratic
terms and the final result is a finite polynomial in the fields $\theta(x)$.
The above imposed technical restrictions on the range of $\theta(x)$ can now 
be lifted.
\\

Several remarks on the result (\ref{equivalence}) are in order: 
When setting all external fields
to zero we find that free Wilson fermions are equivalent to the square of 
the partition function of the self avoiding loop model 
\cite{Pr78,RyHe82} with bending rigidity
$1/\sqrt{2}$ and bond weight $[2 + m]^{-1}$. Thus for a trivial 
background field we reproduce the result for free fermions
obtained by Scharnhorst \cite{Sch97}
with a different method. Eq.~(\ref{equivalence}) is a direct generalization
of the trivial case to fermions in a scalar background field.

From an algebraic point of view Eq.~(\ref{equivalence}) is exactly the
expression we had in mind when discussing the general algebraic 
structure of the determinant. 
In two dimensions the fermion determinant has to be a 
polynomial which can at most be quadratic in the external field. The 
feature that the terms in this polynomial are organized according to closed
loops is inherited from the hopping expansion. Actually, for single 
contributions to (\ref{equivalence}) it is even possible to trace their 
emergence from an expansion of the exponential in (\ref{finalhop}). When 
doing so, one finds that the intersection factor $(-1)^{s(l)}$ provides
the mechanism which ensures the cancellation of loops with multiply occupied
links. Eq.~(\ref{8vertexhop}) establishes that this cancellation
mechanism is also independent of the corner weights. 

Finally, it is obvious from (\ref{equivalence}) that in the loop representation
it is straightforward to integrate out the scalar fields. This can be done with
different actions for $\theta$. In the next section we discuss the case
of a simple Gaussian which will produce the Gross-Neveu model
\cite{GrNe74}. We remark, that
a loop representation for the Gross-Neveu model with staggered fermions has 
been analyzed in \cite{Mo90} and a numerical study of the model with 
conventional methods (introduction of the auxialiary field) is given in 
\cite{CoElRa83}.

\section{Application of the result to the Gross-Neveu model}
\noindent
In order to give an application of our Eq.~(\ref{equivalence}) we now 
integrate the scalar field with a Gaussian measure
\[
\int d\mu[\theta] \; = \; 
\int \prod_{x \in \Lambda}\frac{d \theta(x)}{\sqrt{2 \pi g}}
\exp\left(- \frac{1}{2 g} \theta(x)^2 \right) \; .
\]
When integrating $(\det M[\theta])^N$ with this measure we generate a lattice
version of the $N$-component Gross-Neveu model \cite{GrNe74} 
with (continuum) action
\begin{equation}
S \; = \; \int d^2 x \Big( 
\overline{\psi}(x) \Big[ \gamma_\mu \partial_\mu \; - \; m \Big] \psi(x) \; -
\; g^2 \Big[ \overline{\psi}(x) \psi(x) \Big]^2 \Big) \; ,
\label{grossneveu}
\end{equation}
where $\overline{\psi}, \psi$ now have $N$ flavor components. Using 
(\ref{equivalence}) the lattice partition function for the Gross-Neveu model 
reads
\begin{eqnarray}
Z_{gn} & = & \int d\mu[\theta] \prod_{x \in \Lambda} 
[2\!+\!m\!+\!\theta(x)]^{2N} \;
\left( \sum_{l \in {\cal L}_{sa}}\!\Big( \frac{1}{\sqrt{2}} \Big)^{c(l)}
\!\prod_{y \in P(l)} \frac{1}{2\!+\!m\!+\!\theta(y)} \right)^{2N} 
\nonumber \\
& = & \int d\mu[\theta]
\sum_{l_1, ..., l_{2N}} \Big( \frac{1}{\sqrt{2}} \Big)^{c(l_1) + ...
+ c(l_{2N})} \prod_{x \in \Lambda} [2\!+\!m\!+\!\theta(x)]^{2N - 
O_x(l_1, ..., l_{2N})}
\nonumber \\
 & = & \sum_{l_1, ..., l_{2N}} 
\Big( \frac{1}{\sqrt{2}} \Big)^{c(l_1) + ...
+ c(l_{2N})} \prod_{x \in \Lambda} G[2N - O_x(l_1, ... ,l_N)] \; .
\label{loopgn}
\end{eqnarray}
The function $O_x(l_1, ..., l_{2N})$ counts how many of the independent 
loops $l_1, ..., l_{2N}$ occupy the site $x$ for a given loop configuration.
Each loop can either leave the site empty or occupy it once (the loops 
are self-avoiding) and thus $O_x$ has values between 0 and $2N$.
The function $G[J], J = 0, 1, ..., 2N$ is given by
\[
G[J] \; = \; \int\!\frac{d\theta}{\sqrt{2\pi g}}
e^{- \frac{1}{2g} \theta^2} [2\!+\!m\!+\!\theta ]^J \; = \; 
\sum_{k=0}^{[J/2]} {J \choose 2k} (2k-1)!! \; g^k \; (2+m)^{J-2k}.
\]
Equation (\ref{loopgn}) establishes that the $N$-component Gross-Neveu
model is equivalent to a model of $2N$ independent self-avoiding loops. 
The partition function is a sum over all loop configurations with a weight
corresponding to the total number of corners times a simple function of 
the occupation number for each site. 

The partition function can easily be reformulated as a $8^{2N}$-vertex model.
The new tiles are obtained by using $2N$ different colors and with 
each color one of the line patterns of Fig.~\ref{tiles} is drawn onto our new
tile, giving a total of $8^{2N}$ different tiles. The weight for the 
tiles is given
by a product of the weights $w_i$ for each color (1 for $w_1,w_3,w_4$,
0 for $w_2$ and $1/\sqrt{2}$ for $w_5, ..., w_8$) times the occupation factor
$G[J]$. 
\\

Let's discuss the case of $N=1$ in more detail. Here the tiles show lines
in two colors, say red and blue. The weight factors are the products of 
the $w_i$ for the blue and red lines multiplied by 1 when the tile is empty,
by a factor of $2 + m$ when there is only one color and a factor of 
$(2 + m)^2 + g$ when both colors are used on the tile.
When setting $g = 0$ the weights can be factorized into two terms
corresponding to the two colors and we recover the result for the free
case already discussed above. 

A second choice of parameters also leads to a particularly simple model.
When setting $2+m = 0$, we find that all tiles which show only 
one color vanish. Thus for a non-vanishing contribution 
every site has to either be empty or is visited by both, a blue and
a red loop. For our simple rectangular
lattice this implies,
that the red and blue loop configurations have to sit on top of each other
and integrating out the scalar field at $2+ m = 0$ has produced
a Dirac delta on the space of loops. The resulting model is again a 
self avoiding loop model, now with bending rigidity  $1/2$ and bond weight
$g$. This is the same model which was shown \cite{Sa91}
to be equivalent to the strong coupling Schwinger model. It has been analyzed 
with Monte-Carlo methods in $\cite{GaLaSa92}$ where the existence of 
a phase transition at 
$g_c = 0.5792$ was established. For a study of this model, using
computer algebra on small lattices see \cite{KaMeTu95}.
In the form of the above discussed 
two-color (64-vertex) model, the $N = 1$ Gross-Neveu model is now
relatively simple to analyze numerically in the whole $g,m$-plane. In
particular, the representation as a vertex model may allow for the use of 
efficient cluster algorithms \cite{EvLaMa93}.

\section{Summary and discussion}
\noindent
The motivation for this article was to find simple representations for the
Wilson lattice fermion determinant in an external field. Counting the powers of
the external field in the determinant one finds that a large set of 
contributions appearing in the standard hopping expansion has to cancel. 
The determinant can only be a finite polynomial in the external variables.
Finding, however, a reasonably simple and useful expression for the 
determinant is a hard problem.

In this article we succeeded in finding such a simple representation for the 
case of 2-D Wilson fermions in a scalar background field. The determinant
can be written as the product of two self-avoiding loop models coupled
to the external field. To prove this result, Grassmann techniques and 
hopping expansion for a generalized 8-vertex model were used. In the obtained
loop representation it is straightforward to integrate out the scalar fields,
and the application to the case of the Gross-Neveu model was discussed 
in more detail.

When one instance of a simple expression for a Wilson type 
fermion determinant in
an external field can be found it is natural to ask whether other and more
realistic models also allow for a simple representation of the fermion 
determinant. 

The next candidate are 2-D fermions interacting with an abelian vector field.
The vector field is coupled to the fermions through its gauge
transporters which are supported on links instead of sites. 
This does not pose a fundamental 
problem since the  generalized 8-vertex model 
\cite{Ga98b} can also be formulated for external field living on links.
All terms of the hopping expansion can still be evaluated explicitly.
A certain difficulty is, however, given by the fact that the link variables 
have to be complex conjugated when hopping in
backward direction. Thus the loops in the hopping expansion give contributions
complex conjugated to each other when the orientation is reversed. We 
believe, however, that when properly organizing the loops this problem 
can be overcome. A hint in this direction is Scharnhorst's proof
for the existence of a 
vertex model for the 2-D Thirring model \cite{Sch97}. If a loop 
representation for the determinant in an external vector field exists then
it should be possible to obtain Scharnhorst's result by integrating out the
vector field, similar to the loop representation of the 
Gross-Neveu model which was obtained by 
integrating out the scalar field. The case of non-abelian gauge fields
poses the additional difficulty, that due to the traces over the gauge 
field matrices the product of two loops cannot be written as a new loop. 
Nevertheless, the power counting argument discussed in the introduction 
still suggests, that representations simpler than the standard hopping 
expansion should exist. Similar in 4 dimensions where at least the 
problem of computing the trace of the 4-D
$\gamma$-matrices is 
solved \cite{St81}. 
\\  
\\
{\sl Acknowledgement:} The Author thanks Christian Lang, Klaus Scharnhorst 
and Uwe-Jens Wiese for discussions and remarks on the literature.

\end{document}